\documentclass[12pt]{article}
\usepackage[dvips]{color}
\usepackage{epsfig}
\usepackage{amsmath}
\usepackage{graphicx}
\begin{document}
\title{\bf Bouncing universe with the non-minimally coupled scaler field and its reconstructing}
\author{J. Sadeghi $^{a,}$\thanks{Email: pouriya@ipm.ir}\hspace{1mm} , F. Milani $^{a,}$\thanks{Email: f.m.1683@hotmail.com}
and A. R. Amani$^{b,}$ \thanks{Email: a.r.amani@iauamol.ac.ir}\\
$^a${\small {\em Sciences Faculty, Department of Physics, Mazandaran
University,}}\\{\small {\em P .O .Box 47415-416, Babolsar, Iran}}\\
$^b$ {\small {\em  Department of Physics, Islamic Azad University - Ayatollah Amoli Branch,}}\\
{\small {\em P.O.Box 678, Amol, Iran}}} \maketitle

\begin{abstract}
\noindent \hspace{0.35cm} In this paper we consider a non-minimally coupled scaler field,
 and show its equation of state parameter can crossing over $-1$, $\omega\rightarrow -1$,
 and bouncing condition. Also we obtain the stability conditions and consider reconstructing
 for our model.\\


\end{abstract}
\section{Introduction}
Scalar fields play a central role in the modern cosmology. The
combined analysis of the type Ia supernovae, galaxy clusters
measurements and WMAP data provides an evidence for the accelerated
cosmic expansion {\cite{c1}}-\cite{c5}. The cosmological
acceleration strongly indicates that the present day universe is
dominated by smoothly distributed slowly varying Dark Energy (DE)
component. The modern constraints on the DE state parameter are
around the cosmological constant value, $\omega = -1 \pm 0.1$
{\cite{c3}}-{\cite{c7}} and a possibility that $\omega$ is varied in
time is not excluded. From the theoretical point of view there are
three essentially different cases: $\omega>-1$ (quintessence),
$\omega = -1$ (cosmological constant) and $\omega <
-1$ (phantom) ({\cite{c8}}-{\cite{c27}} and Refs. therein).\\
Since from the observational point of view there is no barrier
between these three possibilities it is worth to consider models
where these three cases are realized. Under general assumptions it
is proved in \cite{c28} that within one scalar field model one can
realize only one possibility: $\omega \geq 1$ (usual model), or
$\omega \leq1$ (phantom model). It is interesting that the
interaction with the cold dark matter does not change the situation
and does not
remove the cosmological constant barrier \cite{c29}.\\
On the other hand, the Friedman equation forms the starting point
for almost all investigations in cosmology. Over the past few years
possible corrections to the Friedman equation have been derived or
proposed in a number of different contexts, generally inspired by
braneworld investigation \cite{c30, c31}. These modification are
often of a form that involves the total energy density $\rho$. In
\cite{c32}, multi-scalar coupled to gravity is studied in the
context of conventional Friedman cosmology. It is found that the
cosmological trajectories can be viewed as geodesic motion in an
augmented target
space.\\
In this way, there are several phenomenological models describing
the crossing of the cosmological constant barrier
\cite{c33}-\cite{c42}. Most of them use more then one scalar field
or use a non-minimal coupling with the gravity, or modified gravity,
in particular via the brane-world scenarios. In two-field models one
of these two fields is a phantom, other one is a usual field and the
interaction is nonpolynomial in general. It is important to find a
model which follows from the fundamental principles and describes a
crossing of
the $\omega = -1$ barrier.\\
In this paper we show that such a model may appear within a brane
approach when the universe is considered as a slowly decaying
D3-brane and a possibility to cross the barrier comes from taking
into account a back reaction of the D3-brane. This DE model
\cite{c23} assumes that our universe is a slowly decaying D3-brane
and its dynamics is described by the open string tachyon mode and
the back reaction of this brane is incorporated in the dynamics of
the closed string tachyon. The open string tachyon dynamics is
described within a level truncated open string field theory (OSFT).
The notable feature of this OSFT description of the tachyon dynamics
is a non-local polynomial interaction \cite{c33}-\cite{c46}.\\
It turns out the open string tachyon behavior is effectively
described by a scalar field with a negative kinetic term (phantom)
\cite{c47}-\cite{c50}. However this model does not suffer from
quantum instability, which usually phantom models have, since in the
nonlocal theory obtained from OSFT there are no ghosts at all near
the non-perturbative vacuum \cite{c23}.\\
On the other hand, inflation \cite{c51}-\cite{c53} is possibly the
only known mechanism which dynamically solves the flatness and the
horizon problem of the universe. Thus it has become an almost
indispensable ingredient in cosmology. The inflaton, a scalar field,
can also produce the density perturbations causally which can match
with the data from observation. For example, the recent WMAP data
\cite{c54}-\cite{c57} strongly supports the idea that the early
universe went through an inflationary phase. Usually one considers
the inflationary phase to be driven by the potential of a scalar
field. Recently there has been an upsurge in activity for
constructing such models in string theory. In the context of string
theory, the tachyon field in the world volume theory of the open
string stretched between a D-brane and an anti-D-brane or on a
non-BPS D-brane has been taken as a natural candidate to play the
role of the inflaton \cite{c58, c59}. This possibility of the
tachyon field driving the cosmological inflation is related to the
decay of unstable brane as a time dependent process which was
advocated by Sen \cite{c60, c61}. The effective action used in the
study of tachyon cosmology consists of the standard Einstein-Hilbert
action and an effective action for the tachyon
field on unstable D-brane or braneantibrane system.\\
In this way, there are a lot of cosmological observations, such as
SNe Ia \cite{c62}-\cite{c65}, WMAP \cite{c66, c67}, SDSS \cite{c68,
c69, c70}, Chandra X-ray Observatory \cite{c71} etc., that they
reveal some cross-checked information of our universe. They suggest
that the universe is spatially flat, and consists of approximately
$70\%$ dark energy with negative pressure, $30\%$ dust matter (cold
dark matters plus baryons), and negligible radiation,
and that the universe is undergoing an accelerated expansion.\\
To accelerate the expansion, the equation of state parameter
$\omega\equiv\frac{p}{\rho}$ of the dark energy must satisfy $\omega
< -1/3$, where $p$ and $\rho$ are its pressure and energy density,
respectively. The simplest candidate of the dark energy is a tiny
positive time-independent cosmological constant $\Lambda$, for which
$\omega = -1$. Another possibility is quintessence
\cite{c72}-\cite{c76}, a cosmic real scalar field that is displaced
from the minimum of its
potential.\\
Briefly, in section 2 we have defined our extended model and we have
shown the equation of state parameter must be crossing over $-1$,
$\omega\rightarrow -1$, and the Hubble parameter $H$ running across
zero at $t = 0$ and ingratiates bouncing conditions. In section 3 we
consider stability of our model. In section 4 we will reconstruct a
non-minimally scaler filed in the three forms of
parametrization. Finally in section 5 we compare them together, with numerical methods.\\

\section{Non-minimally coupled scalar field}

In this section we consider the action in the Jordan frame \cite{c77, c78} with $g_{\mu\nu}$ metric:\\
\begin{eqnarray}\label{ac}
S=\int{d^{4}x\sqrt{-g}\left[\frac{M^{2}_{p}}{2}R-f(\phi)R-\frac{1}{2}(\nabla\phi)^{2}-V(\phi)\right]},
\end{eqnarray}
where $M^{2}_{p}\equiv\frac{1}{8\pi G}$, and $f(\phi)R$ term
corresponds to the non-minimal coupling of the scalar field to
gravity. For a flat Friedman-Robertson-Walker (FRW) universe we can
assume that the universe is described by the flat, homogeneous, and
isotropic universe model with the scale factor $a$. With (1) we
obtain the equation of motion of the scalar field $\phi$,
\begin{eqnarray}\label{pdott}
\ddot{\phi}+3H\dot{\phi}+V'+6f'(\dot{H}+2H^{2})=0.
\end{eqnarray}
Where $V'=\frac{dV(\phi)}{d\phi}$ and $f'=\frac{df(\phi)}{d\phi}$,
and $H=\frac{\dot{a}}{a}$ is the Hubble parameter. The
energy-momentum tensor $T^{\mu\nu}$ is given by the standard:
\begin{eqnarray}\label{met}
\delta_{g_{\mu\nu}}S=-\int{d^{4}x\frac{\sqrt{-g}}{2}T^{\mu\nu}\delta
g_{\mu\nu}}.
\end{eqnarray}
We can read the energy density from (\ref{met}) as
\begin{eqnarray}\label{rho}
\rho=\frac{1}{2}\dot{\phi}^{2}+V+6H(\dot{f}+Hf),
\end{eqnarray}
and similarly the energy pressure
\begin{eqnarray}\label{p}
p=\frac{1}{2}\dot{\phi}^{2}-V-2\ddot{f}-4H\dot{f}-2f(2\dot{H}+3H^{2}).
\end{eqnarray}
With the Friedman equations,
\begin{eqnarray}\label{frid1}
H^{2}=\frac{1}{3M^{2}_{p}}\rho,
\end{eqnarray}
and
\begin{eqnarray}\label{frid2}
\dot{H}=-\frac{1}{2M^{2}_{p}}(\rho+p),
\end{eqnarray}
The continuity equation to be,
\begin{eqnarray}\label{rhodot}
\dot{\rho}+3H\rho(1+\omega)=0,
\end{eqnarray}
and from Eqs. (\ref{rho}) and (\ref{frid1})we have
\begin{eqnarray}\label{h}
H=\frac{1}{6}\frac{6\dot{f}-\sqrt{12\left(\dot{f}(3\dot{f}-\dot{\phi}^{2})+V(M^{2}_{p}-2f)
\right)+6M^{2}_{p}\dot{\phi}^{2}}}{M^{2}_{p}-2f}.
\end{eqnarray}
We now study the cosmological evolution of equation of state for the
present model. The equation of state is $p=\omega\rho$. To explore
the possibility of the $\omega$ across $-1$, we have to check
$\frac{d}{dt}(\rho+p)\neq 0$ when $\omega\rightarrow -1$. From Eqs.
(\ref{rho} ) and (\ref{p}) one can obtain the following expressions,
\begin{eqnarray}\label{rhop}
\rho+p=\dot{\phi}^{2}+2H\dot{f}-4\dot{H}f-2\ddot{f}.
\end{eqnarray}
Therefore from Eqs. (\ref{frid2}) and (\ref{rhop}) we obtain
\begin{eqnarray}\label{hdot}
\dot{H}=-\frac{1}{2(M^{2}_{p}-2f)}(\dot{\phi}^{2}+2H\dot{f}-2\ddot{f}),
\end{eqnarray}
so we have $\frac{d}{dt}(\rho+p)= -2M^{2}_{p}\ddot{H}$ or,
\begin{eqnarray}\label{hddot}
\ddot{H}=\frac{1}{(M^{2}_{P}-2f)}(\frac{d\ddot{f}}{dt}+\dot{H}\dot{f}-H\ddot{f}-\dot{\phi}\ddot{\phi}).
\end{eqnarray}
Now  with replacing Eqs. (\ref{h}) and (\ref{hdot}) in the Eq.
(\ref{hddot}), if $f=1+\sum_{i=1}c_{i}\phi^{2i}$ and a tachyon
scaler field, $V=V_{0}e^{-\lambda\phi^{2}}$ (motivated by string
theory \cite{c79}), with $c_{1}=\frac{1}{12}$ and $c_{i}=0$ for
$i>1$, one can obtain
\begin{eqnarray}\label{hddot1}
\ddot{H}=\frac{1}{(6M^{2}_{p}-\phi^{2})}&\textbf{\{}&\frac{d\ddot{\phi}}{dt}\textbf{[}\phi-\frac{6M^{2}_{p}\dot{\phi}}
{U}\textbf{]}\nonumber\\
&+&\ddot{\phi}\textbf{[}3\dot{\phi}+\frac{6M^{2}_{p}(\lambda\phi
V-2)+2\phi
V(1-\lambda\phi^{2})}{U}+\frac{6M^{2}_{p}\dot{\phi}V(\lambda(\phi^{3}-6M^{2}_{p})-\phi)}{U^{3}}\nonumber\\
&+&\frac{2}{(6M^{2}_{p}-\phi^{2})}\left(\frac{(2\phi
U-3M^{2}_{p}\dot{\phi})\phi\dot{\phi}}{U}
+(\phi\dot{\phi}-U)\phi\right)\textbf{]}\nonumber\\
&+&\dot{\phi}\textbf{[}\frac{2V}{U}\left(2\lambda(3M^{2}_{p}-4\lambda\phi^{2})M^{2}_{p}+\dot{\phi}
(1-\lambda\phi^{2}(5-\lambda\phi^{2}))\right)+\frac{\dot{\phi}}{(6M^{2}_{p}-\phi^{2})}\nonumber\\
&\times&\left(\frac{\phi^{2}}{U}(4U+2V(1+\lambda(6M^{2}_{p}-\phi^{2})))+2(\phi\dot{\phi}-U)
(\frac{4\phi^{2}}{(6M^{2}_{p}-\phi^{2})}+1)\right)\textbf{]}\nonumber\\
&+&\frac{V^{2}(\lambda(\phi^{3}-6M^{2}_{p})-\phi)^{2}+9M^{4}_{p}\dot{\phi}^{2}\ddot{\phi}^{2}}{U}\textbf{\}}.
\end{eqnarray}
Where $U=\sqrt{2(6M^{2}_{p}-\phi^{2})V+6M^{2}_{p}\dot{\phi}}$. With
implying Eqs. (\ref{rhop}) and (\ref{hdot}) and a lot of complexity
calculations one can finds either (i)$\dot{\phi}=0$ or
(ii)$\ddot{\phi}=0$ when $\omega\rightarrow-1$.\\
(i)\hspace{.5cm}Let us assume $\dot{\phi}=0$ first when
$\omega\rightarrow-1$, so we have,
\begin{eqnarray}\label{hddot2}
\ddot{H}&=&\frac{1}{(6M^{2}_{p}-\phi^{2})}\{\phi\frac{d\ddot{\phi}}{dt}+\frac{\ddot{\phi}}{\sqrt{2(6M^{2}_{p}-\phi^{2})V}}
\left(6M^{2}_{p}(\lambda\phi V-2)-2\phi V(1+\lambda\phi^{2})
\right)\nonumber\\
&+&V^{2}(\lambda(\phi^{3}-6M^{2}_{p})-\phi)^{2}\}
\end{eqnarray}
Therefore the conditions for having the $\omega$ across over $-1$
are (i-1)$\frac{d\ddot{\phi}}{dt}\neq 0$, (i-2)$\ddot{\phi}\neq 0$
and (i-3)$\phi^{2}\neq 6M^{2}_{p}$ in addition to the $\dot{\phi}=0$.\\
(ii)\hspace{.5cm}Let us turn to the second case, when
$\omega\rightarrow -1$. We now have
\begin{eqnarray}\label{hddot3}
\ddot{H}=\frac{1}{(6M^{2}_{p}-\phi^{2})}&\textbf{\{}&\frac{d\ddot{\phi}}{dt}\textbf{[}\phi-\frac{6M^{2}_{p}\dot{\phi}}
{U}\textbf{]}\nonumber\\
&+&\dot{\phi}\textbf{[}\frac{2V}{U}\left(2\lambda(3M^{2}_{p}-4\lambda\phi^{2})M^{2}_{p}+\dot{\phi}
(1-\lambda\phi^{2}(5-\lambda\phi^{2}))\right)+\frac{\dot{\phi}}{(6M^{2}_{p}-\phi^{2})}\nonumber\\
&\times&\left(\frac{\phi^{2}}{U}(4U+2V(1+\lambda(6M^{2}_{p}-\phi^{2})))+2(\phi\dot{\phi}-U)
(\frac{4\phi^{2}}{(6M^{2}_{p}-\phi^{2})}+1)\right)\textbf{]}\nonumber\\
&+&\frac{V^{2}(\lambda(\phi^{3}-6M^{2}_{p})-\phi)^{2}}{U}\textbf{\}}.
\end{eqnarray}
Therefore the conditions for having the $\omega$ across over $-1$
are (ii-1)$\frac{d\ddot{\phi}}{dt}\neq 0$, (ii-2)$\dot{\phi}\neq 0$
and (ii-3)$\phi\neq 6M^{2}_{p}$ in addition to the
$\ddot{\phi}=0$.\\
In the Fig. 1 we consider a tachyon scaler field,
$V(\phi)=V_{0}e^{-\lambda \phi^{2}}$ and
$f(\phi)=1+\sum_{i=1}c_{i}\phi^{2i}$. We have shown
$\omega\rightarrow -1$, which gives rise to a possible inflationary
phase after the bouncing.

\begin{figure}[th]
\centerline{\epsfig{file=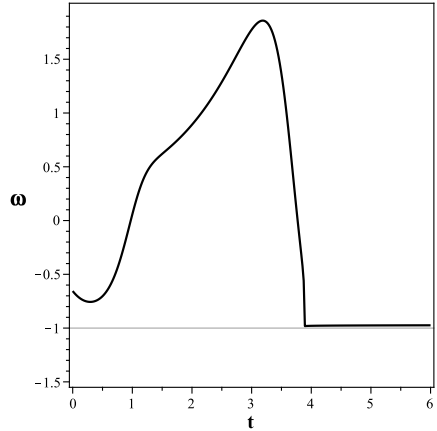,scale=.3}} \caption{ The graph
of EoS for $V_{0}=4$, $\lambda=0.6$, $c_{1}=\frac{1}{12}$ and
$c_{i>1}=0$. Initial values are $\phi(0)=0.5$ and
$\dot{\phi}(0)=-1.55$.}
\end{figure}
Now we consider a detailed examination on the necessary conditions
required for a successful bounce. During the contracting phase, the
scale factor $a(t)$ is decreasing, i.e., $\ddot{a} < 0$, and in the
expanding phase we have $\ddot{a} > 0$. At the bouncing point,
$\ddot{ÿa} = 0$, and around this point $\ddot{a} > 0 $ for a period
of time. Equivalently in the bouncing cosmology the Hubble parameter
$H$ runs across zero from $H < 0$ to $H > 0$ and $H = 0$ at the
bouncing point. A successful bounce requires around this point,
\begin{eqnarray}\label{hdot1}
\dot{H}=-\frac{1}{2M^{2}_{p}}(\rho+p)=-\frac{1}{2M^{2}_{p}}\rho(1+\omega)>0.
\end{eqnarray}
Fig. 2 shows that the Hubble parameter $H$ running across zero at $t
= 0$ and ingratiates bouncing conditions, Eq. (\ref{hdot1}).
\begin{figure}[th]
\centerline{\epsfig{file=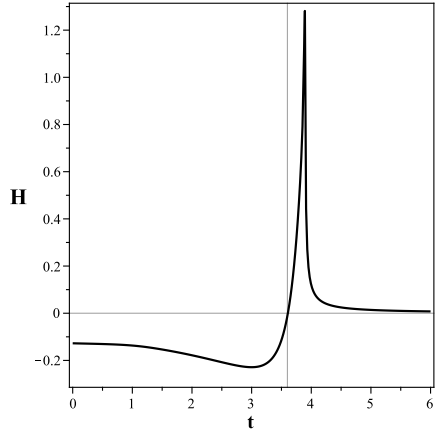,scale=.3}} \caption{ The graph
of Hubble parameter $H$, for the potential $V(\phi)=V_{0}e^{-\lambda
\phi^{2}}$ which $V_{0}=4$, $\lambda=0.6$, $c_{1}=\frac{1}{12}$ and
$c_{i>1}=0$ Initial values are $\phi(0)=0.5$ and
$\dot{\phi}(0)=-1.55$}
\end{figure}
In the Fig. 2 is seen in which the $H$ crossing zero axes and it is
a bounce point.

\section{Stability conditions}
In this section, we deal on the stability of our model. Here we want
to consider the stability by a useful the function $c_s^2=dp/d\rho$.
This function just is the stability of our system by the scalar
field that it must become more than zero. Of course this function
express sound speed in a prefect liquid. Therefore we apply the
analysis of our model in the $\omega' - \omega$ plane which
$\omega'=d\omega/d~ lna$. In that case we obtain from Eqs. (\ref{h})
and as the following form,
\begin{eqnarray}\label{c2}
c^{2}_{s}=\frac{p'}{\rho'}=\omega+\frac{\rho}{\rho'}\omega',
\end{eqnarray}
where the prime denote derivation with respect $ln a$. By
substituting Eq. (\ref{c2}) in term the prime in above equation
yields,
\begin{eqnarray}\label{c21}
c^{2}_{s}=\omega-\frac{\omega'}{3(1+\omega)}.
\end{eqnarray}
To employing the stability condition ($c_s^2>0$), we obtain the two
regions $\omega>-1$ and $\omega<-1$ respectively for
$\omega'>3\omega(1+\omega)$ and $\omega'<3\omega(1+\omega)$ in the
$\omega' - \omega$ phase plane. These regions have been showed in
Fig. 3.
\begin{figure}[th]
\centerline{\epsfig{file=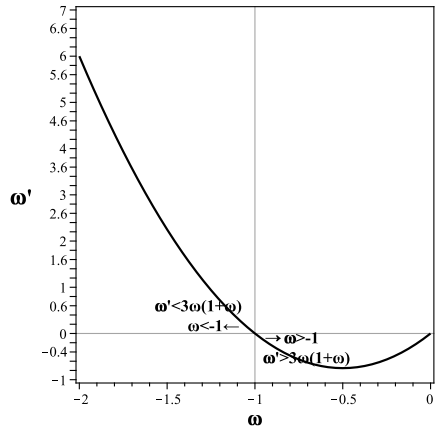,scale=.3}} \caption{ The
graph of $\omega'$ with respect to $\omega$ for the two cases
investible.}
\end{figure}

Now we show the stability condition for non-minimally coupled scalar
field. We consider the stability condition by $c^{2}_{s}$ parameter,
which it can rewrite as the Hubbel parameter,
\begin{eqnarray}\label{cs2}
c^{2}_{s}=-1-\frac{\ddot{H}}{3H\dot{H}}.
\end{eqnarray}
Now by numerical computing we can plot the $c^{2}_{s}$, in term of
time evolution, that is shown in Fig. 4. We can see the stability
condition for late time evolution i.e. $c^{2}_{s}$ for $t\rightarrow
+\infty$.
\begin{figure}[th]
\centerline{\epsfig{file=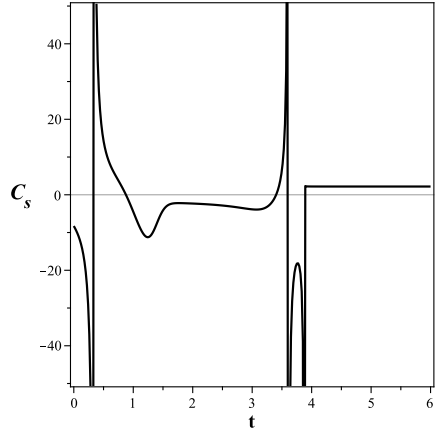,scale=.3}} \caption{ The
graph of $c^{2}_{s}$ with respect to time evolution.}
\end{figure}

\section{Reconstructing a non-minimally coupled scalar field}
In this section we consider reconstruct a non-minimally scaler filed
in the three forms of parametrization. \\
Therefore from Eqs, (\ref{rho}) and (\ref{p}) one can finds,
\begin{eqnarray}\label{2rho}
2\rho+3p=\frac{5M^{2}_{p}}{2(M^{2}_{p}-2f)}\dot{\phi}^{2}-\frac{M^{2}_{p}}{(M^{2}_{p}-2f)}V-\frac{6M^{2}_{p}}{(M^{2}_{p}-2f)}\ddot{f}=-\tilde{K},
\end{eqnarray}
with comparing by Eq. (\ref{rho}) and implying Eq. (\ref{h}) we
have,
\begin{eqnarray}\label{rho-1}
\rho=\tilde{K}+3\tilde{V}.
\end{eqnarray}
Where
\begin{eqnarray}\label{vbar1}
\tilde{V}&=&\frac{M^{2}_{p}}{(M^{2}_{p}-2f)^{2}}\left(2\dot{f}+(\dot{\phi}^{2}-2\ddot{f})(M^{2}_{p}-2f)\right)\nonumber\\
&-&\frac{M^{2}_{p}\dot{f}}{3(M^{2}_{p}-2f)^{2}}\sqrt{36\dot{f}^{2}+6(M^{2}_{p}-2\dot{f})\dot{\phi}^{2}+12V(M^{2}_{p}-2f)}.
\end{eqnarray}
In this method  the energy pressure is,
\begin{eqnarray}\label{p-1}
p=-(\tilde{K}+2\tilde{V}),
\end{eqnarray}
and equation of state can be obtained such as following,
\begin{eqnarray}\label{omega-1}
\omega=-1+\left(\frac{\tilde{V}}{\tilde{K}+3\tilde{V}}\right).
\end{eqnarray}
Then we can rewrite the Friedman equations from Eqs, (\ref{frid1})
and (\ref{frid2}) as following,
\begin{eqnarray}\label{3m}
3M^{2}_{p}H^{2}=\rho_{m}+\rho=\rho_{m}+\tilde{K}+3\tilde{V},
\end{eqnarray}
and
\begin{eqnarray}\label{2m}
2M^{2}_{p}\dot{H}=-(\rho_{m}+\rho+p)=-\rho_{m}-\tilde{V}.
\end{eqnarray}
Where $\rho_{m}$ is the energy density of dust matter, therefore we
have
\begin{eqnarray}\label{kbar}
\tilde{K}=2\rho_{m}+3M^{2}_{p}H^{2}+6M^{2}_{p}\dot{H},
\end{eqnarray}
and
\begin{eqnarray}\label{vbar}
\tilde{V}=-\rho_{m}-2M^{2}_{p}\dot{H}.
\end{eqnarray}
In this model, the dark energy fluid does not couple to the
background fluid, the expression of the energy density of dust
matter in respect of redshift $z$ is \cite{c80},
\begin{eqnarray}\label{rhom}
\rho_{m}=3M^{2}_{p}H^{2}_{0}\Omega_{m0}(1+z)^{3},
\end{eqnarray}
where $\Omega_{m0}$ is the ratio density parameter of matter fluid
and the subscript $0$ indicates the present value of the
corresponding quantity. Using the following relation
\begin{eqnarray}\label{ddt}
\frac{d}{dt}=-H(1+z)\frac{d}{dz},
\end{eqnarray}
one can rewrite $\tilde{K}$ and $\tilde{V}$ as following
\begin{eqnarray}\label{kbar1}
\tilde{K}=M^{2}_{p}H^{2}_{0}\left(6\Omega_{m0}(1+z)^{3}+3r-3(1+z)r^{(1)}\right),
\end{eqnarray}
and
\begin{eqnarray}\label{vbar1}
\tilde{V}=M^{2}_{p}H^{2}_{0}\left(-3\Omega_{m0}(1+z)^{3}+(1+z)r^{(1)}\right).
\end{eqnarray}
Where $r= \frac{H^{2}}{H^{2}_{0}}$ and
$r^{(n)}=\frac{d^{n}r}{dz^{n}}$. By using Eqs. (\ref{omega-1}),
(\ref{kbar1}) and (\ref{vbar1}) we obtain following expression for
equation of state,
\begin{eqnarray}\label{omega-2}
\omega=\frac{(1+z)r^{(1)}-3r}{-3\Omega_{m0}(1+z)^{3}+3r},
\end{eqnarray}
and with using Eqs. (\ref{rho-1}), (\ref{p-1}) and (\ref{ddt}) one
can find the sound speed such as following,
\begin{eqnarray}\label{c-1}
c^{2}_{s}=\frac{(1+z)r^{(2)}-2r^{(1)}}{-9\Omega_{m0}(1+z)^{2}+3r^{(1)}}.
\end{eqnarray}
the sound speed is discussed for investigation of stability of the
model and it necessary is to be $c^{2}_{s}\geq 0$.\\
Finally from Eq. (\ref{omega-2}) we can obtain following equation
for $r(z)$
\begin{eqnarray}\label{r}
r(z)=\Omega_{m0}(1+z)^{3}+(1-\Omega_{m0})e^{3\int^{z}_{0}\frac{1+\omega(\tilde{z})}{1+\tilde{z}}d\tilde{z}}
\end{eqnarray}
Also by using Eq (\ref{vbar1}) we have an expression for
deceleration parameter $q$ as follows,
\begin{eqnarray}\label{q}
q=-1-\frac{\dot{H}}{H^{2}}=\frac{(1+z)r^{(1)}-2r}{2r}.
\end{eqnarray}
\section{Parametrization}
Now act three different forms of parametrization
as following and compare them together, with numerical methods.\\
\textbf{Parametrization 1:}\\
First parametrization has proposed by Chevallier and Polarski
\cite{c81} and Linder \cite{c82}, where the EoS of dark energy in
term of redshift $z$ is given by,
\begin{eqnarray}\label{z1}
\omega(z)=\omega_{0}+\frac{\omega_{a}z}{1+z}.
\end{eqnarray}
\textbf{Parametrization 2:}\\
Another the EoS in term of redshift $z$ has proposed by Jassal,
Bagla and Padmanabhan \cite{c83} as,
\begin{eqnarray}\label{z2}
\omega(z)=\omega_{0}+\frac{\omega_{b}z}{(1+z)^{2}}.
\end{eqnarray}
\textbf{Parametrization 3:} \\
Third parametrization has proposed by Alam, Sahni and Starobinsky
\cite{c84}. They take expression of $r$ in term of $z$ as
followoing,
\begin{eqnarray}\label{z3}
r(z)=\Omega_{m0}(1+z)^{3}+A_{0}+A_{1}(1+z)+A_{2}(1+z)^{2}.
\end{eqnarray}
By using the results of Refs.\cite{c85, c86, c87}, we get
coefficients of parametrization 1 as $\Omega_{m0} =0.29$,
$\omega_{0} = -1.07$ and $\omega_{a} = 0.85$, coefficients of
parametrization 2 as $\Omega_{m0} = 0.28$, $\omega_{0} = -1.37$ and
$\omega_{b} = 3.39$ and coefficients of parametrization 3 as
$\Omega_{m0} = 0.30$,
$A_{0} = 1$, $A_{1} = -0.48$ and $A_{2} = 0.25$.\\
The evolution of $\omega(z)$ and $q(z)$ are plotted in Fig. 5 and
Fig. 6 respectively. Also, using Eqs. (\ref{kbar1}), (\ref{vbar1})
and the three parameterizations, the evolutions of $K (\tilde{z})$
and $V (\tilde{z})$ are shown in Fig. 7 and Fig. 8 respectively.
\begin{figure}[th]
\centerline{\epsfig{file=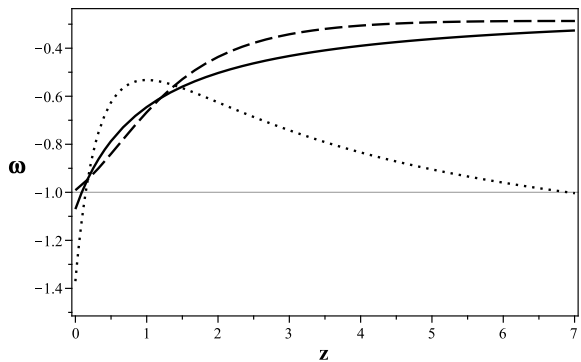,scale=.4}} \caption{ Graphs
for the EoS parameter in respect of redshift $z$. The solid,
 dot and dash lines represent parametrization 1, 2 and 3 respectively.}
\end{figure}
\begin{figure}[th]
\centerline{\epsfig{file=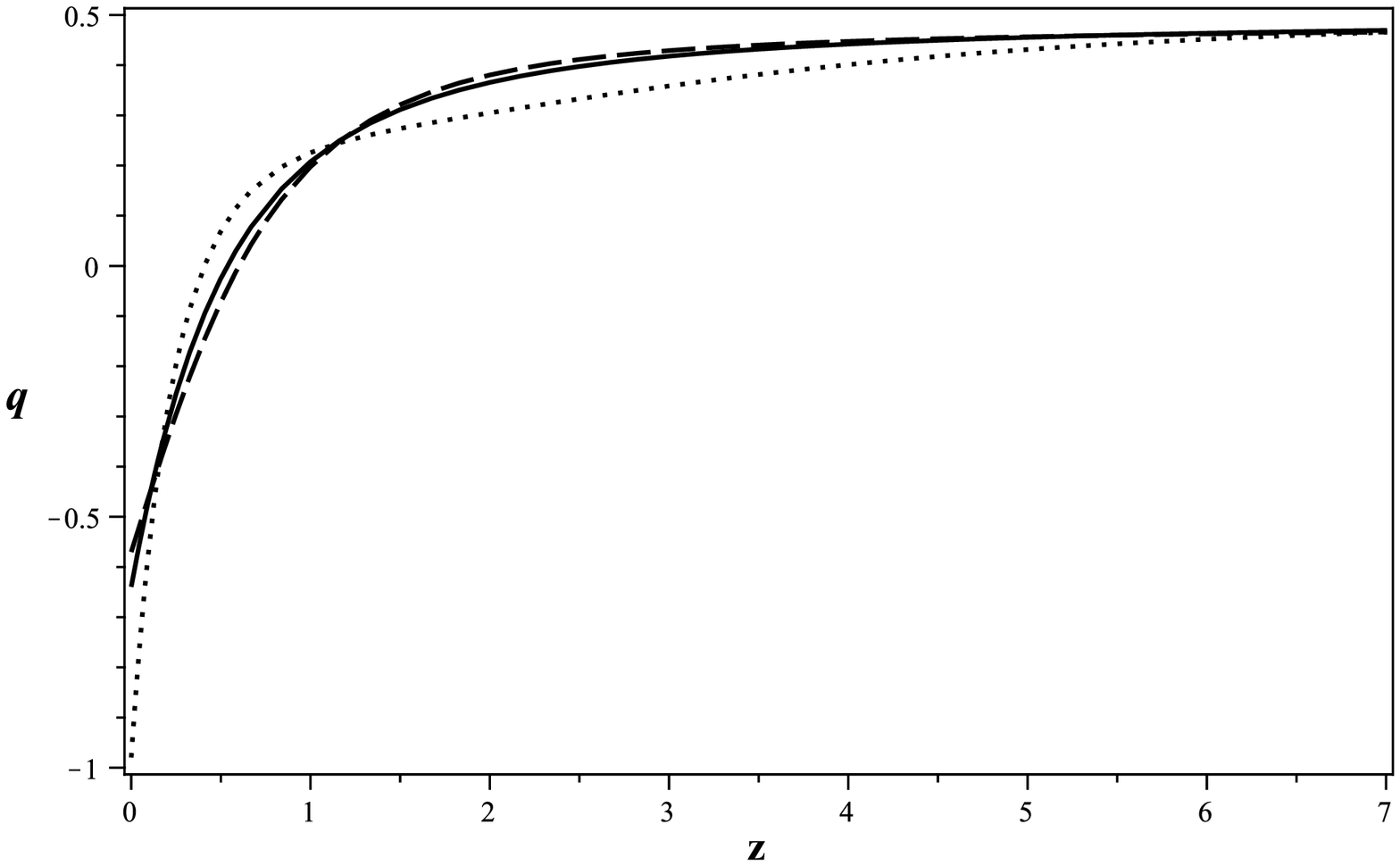,scale=.4}} \caption{ Graphs
for the deceleration parameter in respect of redshift $z$. The
solid, dot and  dash lines represent parametrization 1, 2 and 3
respectively.}
\end{figure}

\begin{figure}[th]
\centerline{\epsfig{file=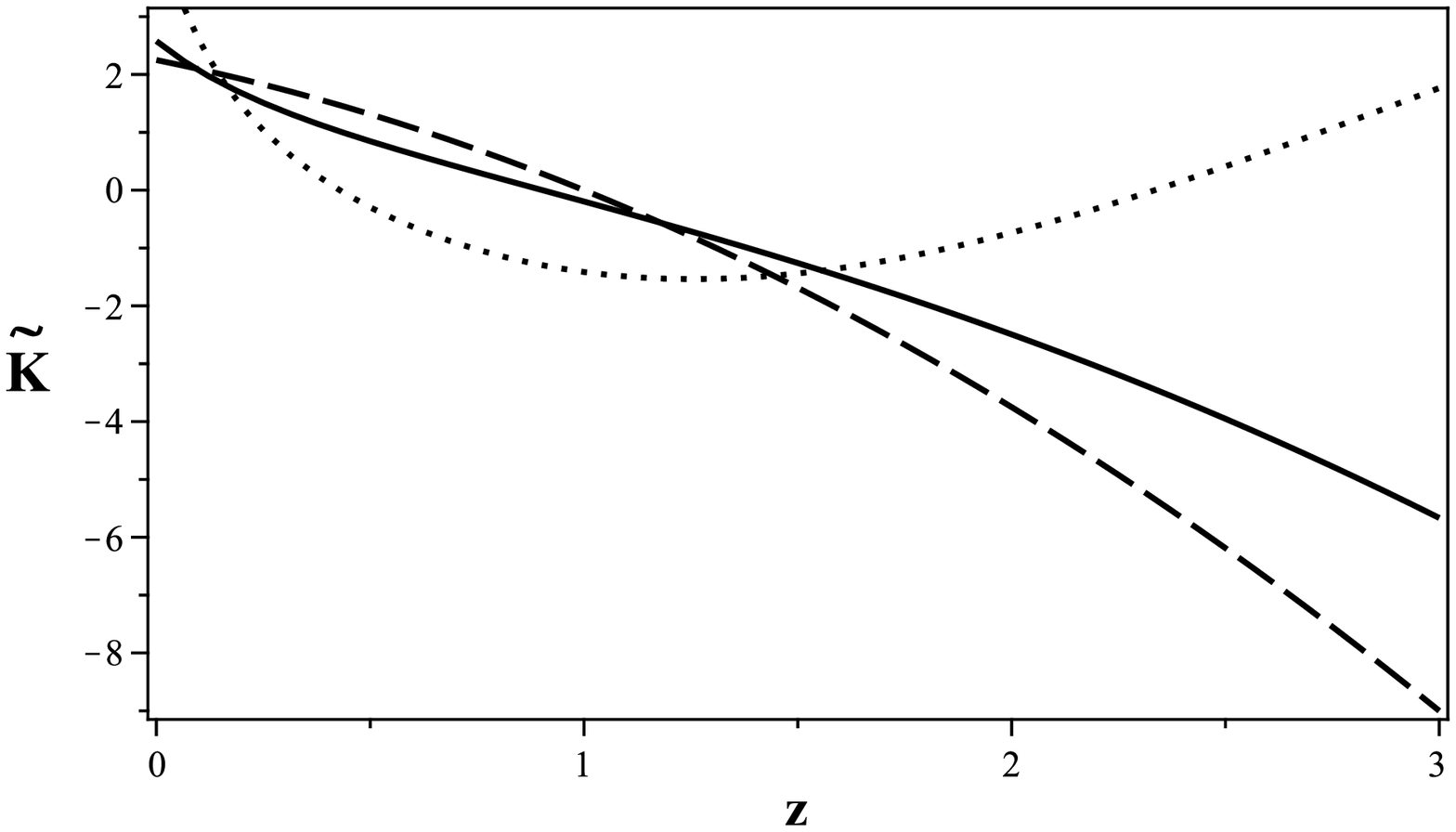,scale=.4}} \caption{ Graphs
for the reconstructed $\tilde{K}$ in respect of redshift $z$. The
solid, dot and dash lines represent parametrization 1, 2 and 3
respectively.}
\end{figure}

\begin{figure}[th]
\centerline{\epsfig{file=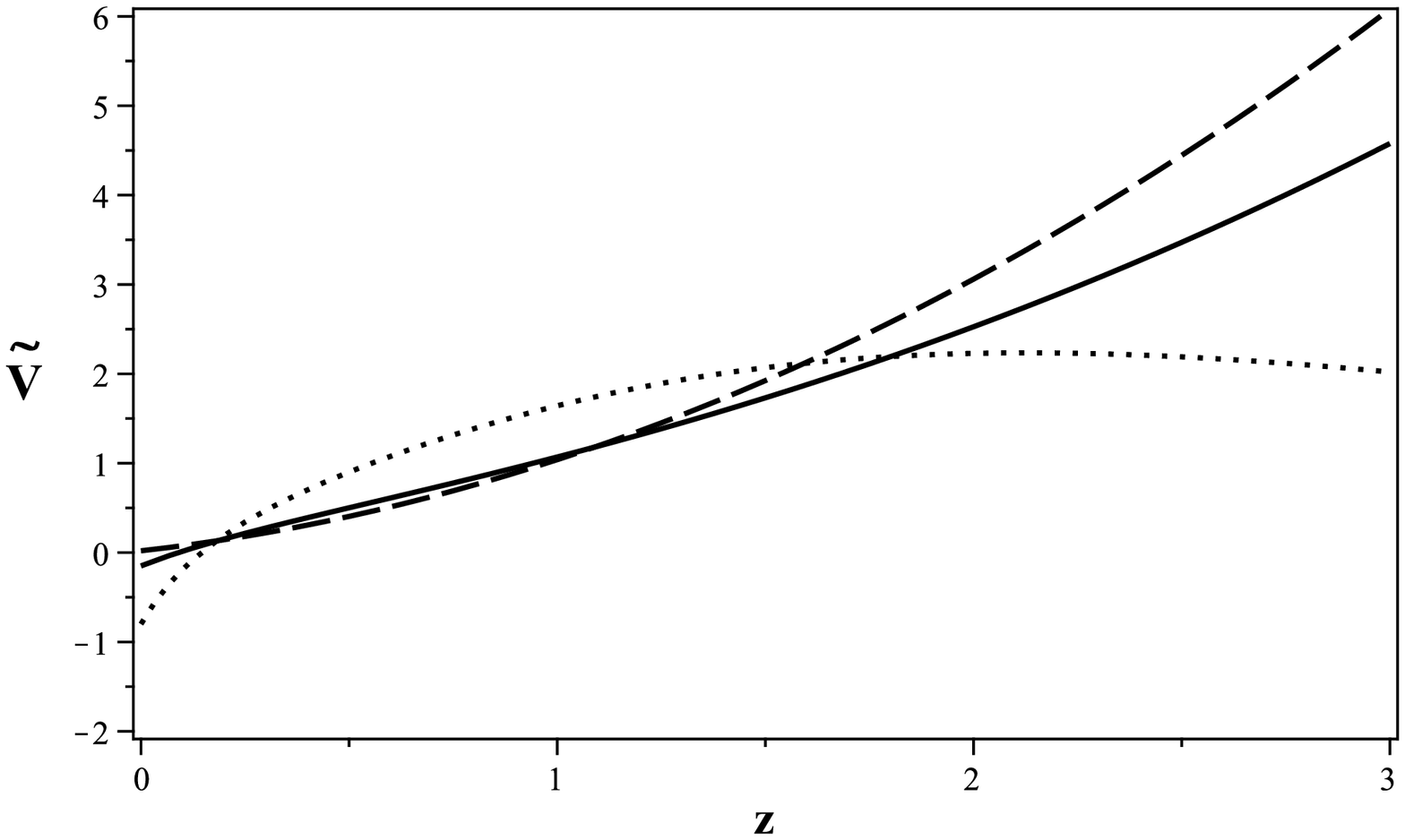,scale=.4}} \caption{ Graphs
for the reconstructed $\tilde{V}$ in respect of redshift $z$. The
solid, dot and  dash lines represent parametrization 1, 2 and 3
respectively.}
\end{figure}
\begin{figure}[th]
\centerline{\epsfig{file=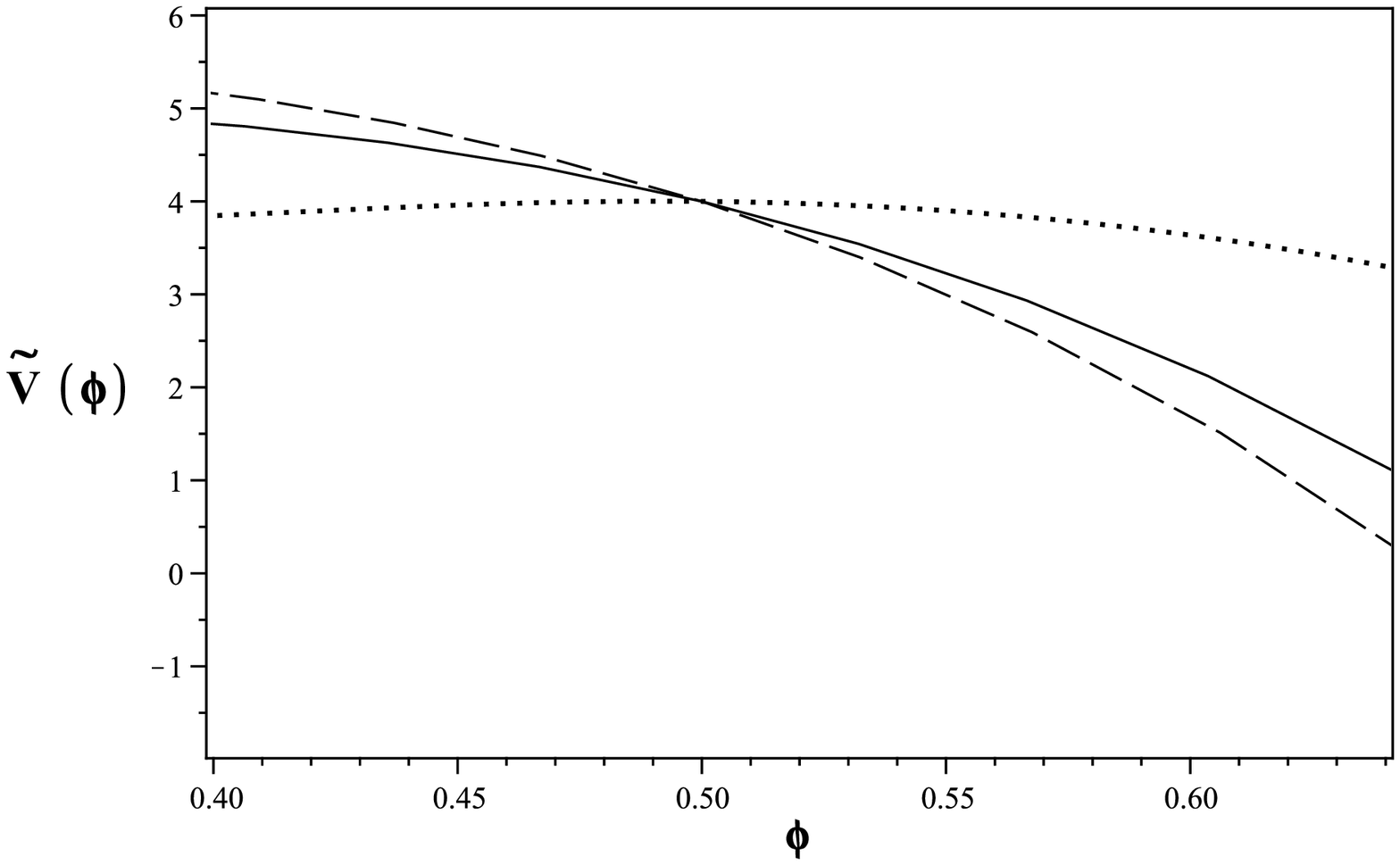,scale=.4}} \caption{
Graphs for the reconstructed $\tilde{V}$ in respect of scaler field
$\phi$. The solid, dot and  dash lines represent parametrization 1,
2 and 3 respectively.}
\end{figure}

The evolution of a scalar field in respect of redshift $z$ are sam
for all of parametrization 1, 2 and 3. therefore all of
parametrization give us consequence well. Fig. 5 show us in which we
can explicitly see the dynamics of a scalar field . Also
slope of graph decrease in the early epoch.\\
Now to achieve to stability of the model by using Eq. (\ref{c-1}),
so we can obtain following condition for all of parametrization
\begin{eqnarray}\label{req3}
r(z)\geq \Omega_{m0} (1+z)^3,
\end{eqnarray}
where is accurate for three above parametrization.

\section{Conclusion}
In this paper , we have investigated bouncing universe by the action
in the Jordan frame metric with Eq. (\ref{ac}). In this form has
coupled a functional of scaler field with gravity term. We obtained
equation of state with respect to time evolution, and one has
crossing $\omega\rightarrow -1$. By consider stability, we have
drown speed of sound in term of time evolution and it has shown
existence of stability in late time. In the graph of Hubble
parameter, we can see
bouncing condition as $H$ crossing from $t=0$ i.e. $\dot{H}>0$.\\
Also, we have reconstructed all of cosmology parameters with respect
to redshift $z$. In that case, we have described the EoS to cross
from $-1$ for three parametrization. Therefore, we can see from
graph of $z$, the second parametrization is better than two other
parametrization.

\end{document}